# Understanding the role of Ca segregation on thermal stability, electrical resistivity and mechanical strength of nanostructured aluminum

*Xavier Sauvage[1*], Fabien Cuvilly[1], Alan Russell[2], Kaveh Edalati[3]*

1- Normandie Université, UNIROUEN, INSA Rouen, CNRS, Groupe de Physique des Matériaux, 76000 Rouen, France

2- Department of Materials Science and Engineering, Iowa State University, Ames, IA 50011, USA

3- WPI, International Institute for Carbon-Neutral Energy Research (WPI-I2CNER), Kyushu University, Fukuoka, Japan

* Corresponding author: xavier.sauvage@univ-rouen.fr

**Abstract:** Achieving a combination of high mechanical strength and high electrical conductivity in low-weight Al alloys requires a full understanding of the relationships between nanoscaled features and physical properties. Grain boundary strengthening through grain size reduction offers some interesting possibilities but is limited by thermal stability issues. Zener pinning by stable nanoscaled particles or grain boundary segregation are well-known strategies for stabilizing grain boundaries. In this study, the Al-Ca system has been selected to investigate the way segregation affects the combination of mechanical strength and electrical resistivity. For this purpose, an Al-Ca composite material was severely deformed by high-pressure torsion to achieve a nanoscaled structure with a mean grain size of only 25 nm. X-ray diffraction, transmission electron microscopy and atom probe tomography data revealed that the fcc Ca phase was dissolved for large levels of plastic deformation leading mainly to Ca segregations along crystalline defects. The resulting microhardness of about 300 HV is much higher than predictions based on Hall and Petch Law and is attributed to limited grain boundary mediated plasticity due to Ca segregation. The electrical resistivity is also much higher than that expected for nanostructured Al. The main contribution comes from Ca segregations that lead to a fraction of electrons reflected or trapped by grain boundaries twice larger than in pure Al. The two-phase state was investigated by in-situ and ex-situ microscopy after annealing at 200°C for 30 min, where precipitation of nanoscaled $Al_4Ca$ particles occurred and the mean grain size reached 35 nm. Annealing also significantly decreased electrical resistivity, but it remained much higher than that of nanostructured pure Al, due to $Al/Al_4Ca$ interfaces that reflect or trap more than 85% of electrons.

## 1- Introduction

The design of low-weight electrical conductors with a high specific mechanical strength combined with a high electrical conductivity is of interest not only for automotive and aircraft industries but also for power transmission lines. Aluminum, with a relatively low production cost, an electrical conductivity of about 62% IACS (International Annealed Copper Standard) and a density of about 2.7 against 8.9 g.cm$^{-3}$ for Cu, is the most interesting candidate [1, 2]. However, like copper, the low mechanical strength of pure Al requires advanced microstructural design to reach high mechanical strength without substantially degrading the electrical properties. Moreover, due to its higher electrical resistivity (leading to more energy dissipation by Joule effect) and relatively low recrystallization temperature, special attention should be given to optimizing the microstructure's thermal stability.

Classical precipitation hardening is sometimes used for power transmission lines using Al-Mg-Si alloys [1], but fine precipitate particles scatter electrons, causing significant decreases in electrical conductivity. It has been demonstrated however that a similar or even a higher mechanical strength could be achieved in ultrafine grain (UFG) or nanoscaled Al [3-7]. This is the well-known grain boundary (GB) strengthening as described by the Hall and Petch Law [6-8]. For example, nanocrystalline sputtered Al films exhibit a yield stress higher than 400 MPa [5]. In such materials with grain size below 100 nm, the uniform elongation is often very much reduced, and some specific deformation mechanisms such as twinning [9-11] and GB sliding [12] or GB migration [13] occur.

UFG structures can be achieved by Severe Plastic Deformation (SPD) [14] with the final grain size being controlled by dynamic recovery and recrystallisation processes. Solute elements are needed to achieve grain sizes in the submicrometer range [15]. Using such a strategy, it has been demonstrated that the combination of GB and precipitation hardening may give rise to optimized combinations of mechanical and electrical properties in 6000-series alloys [16] and to a yield stress up to 1 GPa in other Al alloys [17, 18]. However, residual alloying elements in solid solution reduce the electrical conductivity, and the overall thermal stability is relatively low. The thermal stability of UFG or nanostructured alloys might be improved by nanoscaled particles (Zener pinning) [19] or segregations to reduce the GB energy (and thus the driving force for coarsening) or directly the boundary mobility [20-22].

Thus, the ideal UFG structure for a combination of low weight, high strength, high electrical conductivity and high thermal stability requires a small amount of an immiscible alloying element to achieve pure Al nanoscaled grains pinned by GB segregations or nanoscaled particles. The best strategy might depend on the considered system, but there is little knowledge to optimize the nanoscaled structure regarding the envisioned multifunctional properties. It is well known that the electrical resistivity could be significantly increased by crystalline defects, especially dislocations or GBs [23-26]. Twins, such as those grown during electrodeposition of nanocrystalline Cu, are a kind of exception [27]. However, the influence of segregations, both on electrical resistivity and mechanical strength of nanostructured Al alloys is not yet completely understood. It has already been demonstrated that grain refinement and intermetallic fragmentation can be successfully achieved by SPD processing of as-cast Al alloys containing immiscible elements (Rare Earth [28], Fe [19, 29] or Ca [30]). However, starting from as-cast structures with intermetallic compounds requires huge plastic strains, and only Zener pinning can be achieved. The aim of the present work was to investigate both the influence of segregations and of intermetallic phases in a nanostructured Al-Ca alloy. To reach this goal, it was assumed that the co-deformation of fcc Al and fcc Ca phases could lead to mechanical mixing, promote the nanoscaled structure formation and segregation of the minor element (namely Ca) along grain boundaries, as observed in various immiscible systems [31-36]. Thus, a bi-metallic composite material made of fcc Al and fcc Ca phases [37] has been processed by High-Pressure Torsion (HPT) at various strains. In a second step, intermetallic particles were nucleated in a controlled post-processing manner to preserve the overall nanoscale structure. Microstructures have been characterized down to the nanoscale to correlate with mechanical and electrical properties with a special emphasis on the critical role played by the segregation of Ca atoms at GBs.

## 2- Experimental

The material investigated in the present study is an Al-Ca metal matrix composite produced from high-purity Al and Ca powders that have been mixed, compacted and extruded to achieve a rod of 10mm in diameter with a volume fraction of 12% Ca (Details about the processing route are given in [37]). During the extrusion process, equiaxed Ca particles are transformed into sub-micrometer co-axial filaments. Then, the Al-Ca rod was sliced to prepare discs (diameter 10 mm, thickness 0.8 mm) for further deformation by (HPT) under a pressure of 6 GPa with a speed of one revolution per minute. Samples were deformed by 1, 10, 100 or 1000 revolutions at room temperature. The resulting shear deformation $\gamma$ was estimated as: $\gamma = 2\Pi\, r\, N\, /\, t$, where $r$ is the distance from the disc center, $N$ the number of revolutions and $t$ the disc thickness. Thanks to the large dimension of anvils as compared to the sample volume, the heat generated during the process was efficiently dissipated. The temperature measured with a pyrometer on the anvils near the sample always stayed below 50 °C, even after 1000 revolutions.

The mechanical strength of the severely deformed Al-Ca composite was evaluated by Vickers micro-hardness (HV) measurements carried out across HPT disc diameters with a load of 300 g. The electrical resistivity was evaluated at room temperature using an eddy-current electric conductivity meter calibrated with pure Al. Due to the relatively large zone probed with this technique (about 5 mm in diameter), measurements were performed in the center and near the edge of HPT discs, providing only two average values for each condition.

Crystallographic phases were identified by X-ray diffraction (XRD) using Brucker D8 diffractometer with a Co anticathode (Co-K$\alpha$ radiation, $\lambda$ = 0.17909 nm). Data were recorded with the torsion axis as the out-of-plane component.

Microstructures were characterized by Transmission Electron Microscopy (TEM) on a JEOL ARM 200F microscope. Images were also recorded in Scanning TEM (STEM) mode with a bright field (BF) detector (collection angles 0-45 mrad) and a High-Angle Annular Dark-Field (HAADF) detector (collection angles 68-280 mrad). X-ray Energy Dispersive Spectroscopy (EDS) was carried out with a JEOL Centurio detector (collection angle: 1 sr). TEM samples were prepared in the cross section of HPT discs by Focused Ion Beam (FIB) using a Zeiss N-vision40 dual beam Scanning Electron Microscope (SEM). The thermal stability at 200 °C was checked using an in-situ TEM double-tilt heating holder (Gatan 652 MA). Heating from room temperature to 200 °C was done in a one-minute interval. Based on these observations, the static annealing conditions have been established, and some HPT samples were annealed for 30 min at 200 °C.

More detailed information about the Ca distribution has been obtained by Atom Probe Tomography (APT) analyses. They were carried out with a CAMECA LEAP-4000HR apparatus in ultra-high vacuum ($10^{-11}$ mbar), with a pulse fraction of 20%, a pulse repetition rate of 200 kHz and a sample temperature of 50 K. Three-dimensional (3D) reconstructions were performed using IVAS Software and further data processing was done with the Gpm3dSoft Software. APT tips were prepared by FIB so that the shear plane was perpendicular to the analysis direction.

## 3- Results

To evaluate hardness changes due to SPD of the Al-Ca composite, profiles were measured across HPT discs (Fig. 1). After one revolution, the hardness remains below 50 HV, and there is only a small gradient from 25 HV in the center up to about 35 HV in the outer part. Increasing the number of revolutions (i.e., the shear strain) by a factor of ten leads to significant changes, and while the hardness remains below 50 HV in the center part, it rises to about 250 HV in the outer part. After 100 revolutions, the hardness is more homogeneous across the HPT disc, it seems to saturate between 280 and 300 HV at the periphery while the 1 mm softer region remains in the center. At higher shear strain (up to 1000 revolutions), the hardness is relatively homogeneous across the HPT disc, but a significant softening (about 50 HV) is clearly observed as compared to the sample processed up to 100 revolutions. To evaluate the thermal stability, annealing during 30 min at 200 °C was carried out on the sample processed by 100 revolutions, and Fig. 1 shows that this annealing causes a hardness drop of about 100 HV.

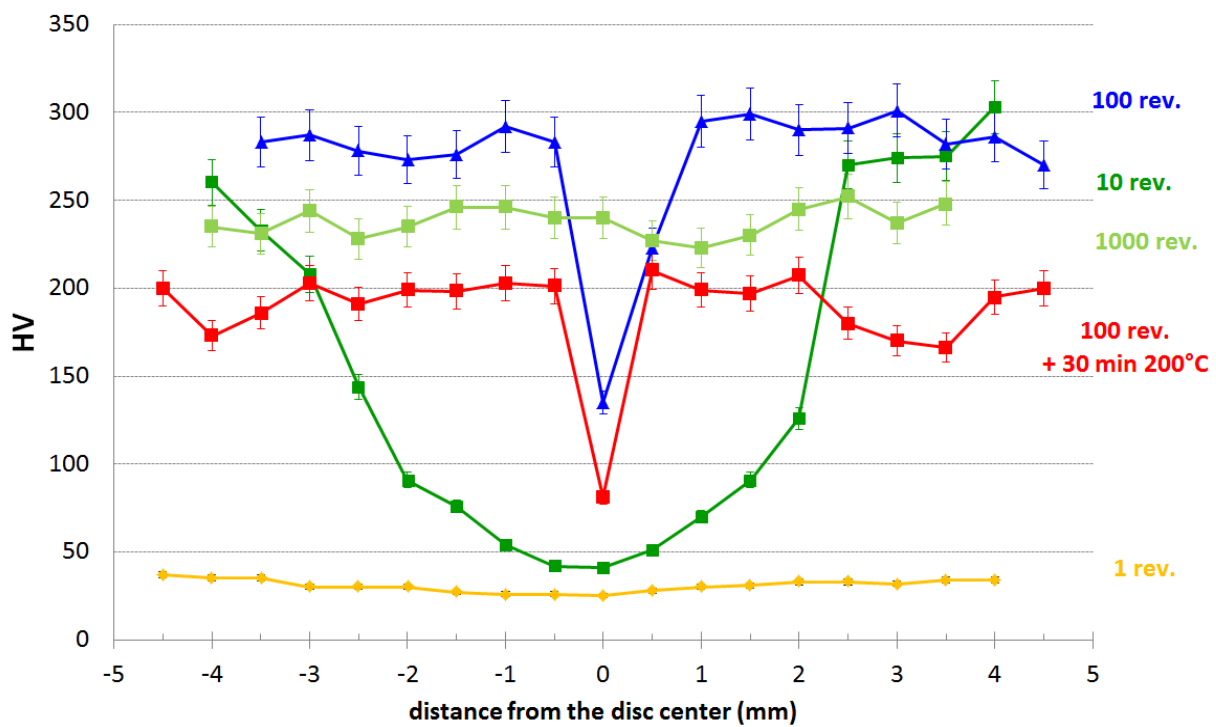

*Figure 1: Micro-hardness profiles measured across the discs processed by HPT (1, 10, 100 and 1000 revolutions) and after post-HPT annealing (100 revolutions followed by 30 min at 200 °C).*

The electrical resistivity and the hardness of the Al-Ca composite processed by HPT are plotted as a function of the shear strain on Fig. 2. At lower strains, up to $\gamma = 100$, they both increase slightly without exceeding 400 $10^{-4} \mu\Omega m$ and 60 HV respectively and thus remain relatively close to the properties of the initial Al-Ca composite material [37]. However, at a shear strain of about $\gamma$ = 200, there is a very sharp increase of both the electrical resistivity and the microhardness that reach a plateau for $\gamma$ > 1000, respectively close to 2100±100 $10^{-4}\mu\Omega$m and 290±10 HV. The short time annealing (30 min at 200°C) leads to a significant drop, down to about 1200±100 $10^{-4}\mu\Omega$m and 200±10 HV, but still much larger than original properties of the Al-Ca composite.

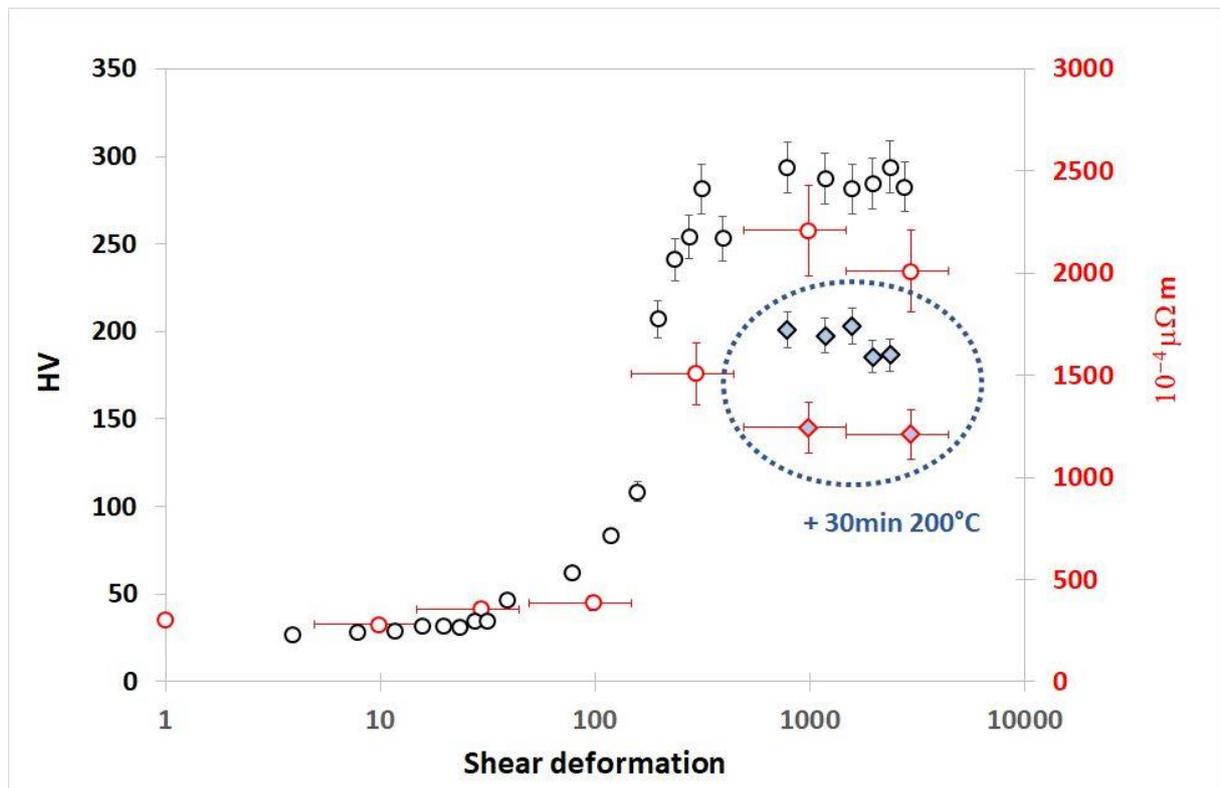

*Figure 2:* Evolution of the microhardness (HV, black markers, left vertical scale) and electrical resistivity ($10^{-4}$ $\mu\Omega$m, red markers, right vertical scale) as a function of the shear deformation induced by HPT. Circle markers are for the as-deformed material and squares for the material aged 30 min at 200 °C. Measurements were carried out on samples processed by 1, 10 and 100 revolutions. The initial electrical resistivity is from [37].

The evolution of phases during SPD has been characterized by XRD. In samples processed by one and 10 revolutions (Fig. 3a), peaks corresponding to fcc Al, fcc Ca, and SiC were detected. The SiC was detected only in the one- and 10-revolution samples, since these samples were very soft (Fig. 1) and ductile. After 100 revolutions, the material became much harder (Fig. 1), and no SiC particles are incorporated during polishing. In the 100-revolutions material, only the fcc Al phase remains (along with significant peak broadening), and no peaks attributable to fcc Ca were detected. This seems to indicate that some mechanical mixing led to the dissolution of the Ca phase during HPT processing. At a larger shear strain, up to 1000 revolutions (Fig. 3b), the spectrum looks similar, but additional peaks corresponding to the $Al_2Ca$ phase are clearly exhibited. After annealing for 30 min at 200 °C, the sample processed by 100 revolutions displayed significant fcc Al peak-width reduction, indicating some defect recovery (Fig. 3b). There are also numerous peaks attributed to the $Al_4Ca$ phase showing that this intermetallic phase has nucleated.

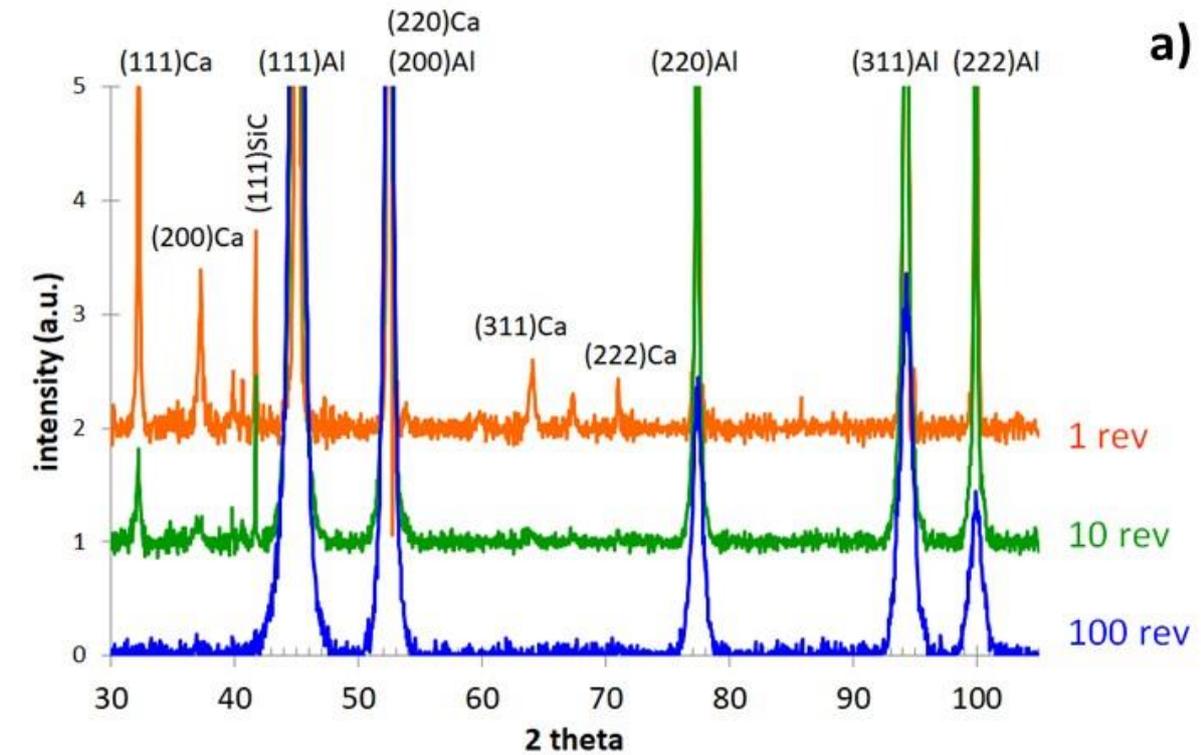

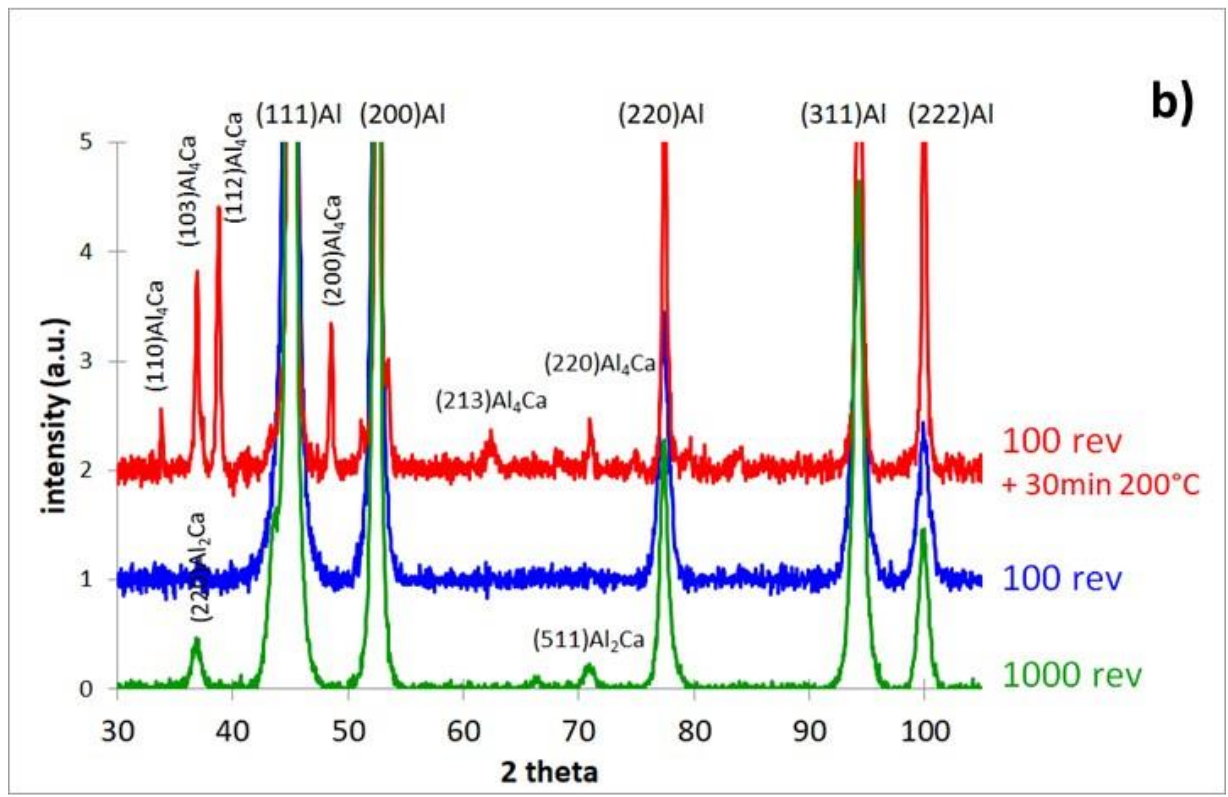

*Figure 3:* (a) XRD patterns showing the progressive dissolution of Ca in fcc Al during deformation up to 100 revolutions by HPT, (b) XRD patterns showing the nucleation of $Al_2Ca$ phase during HPT up to 1000 turns and the precipitation of the $Al_4Ca$ phase in the material deformed up to 100 revolutions and annealed 30 min at 200 °C.

The microstructure resulting from SPD of the Al-Ca composite has been investigated by TEM in the sample processed by 100 revolutions where the maximum hardness was achieved (Fig. 1) and where the fcc Ca phase could not be detected by XRD (Fig. 3a). As shown on the bright-field image of Fig. 4a, the grain size has a mean value between 20 and 30 nm. The corresponding Selected Area Electron Diffraction (SAED) pattern exhibits Debye-Scherrer rings typical of nanostructured fcc Al (Fig. 4b), fully consistent with XRD data (Fig. 3a). However, a few spots that could be indexed with reflections corresponding to the interplanar distances of the $Al_4Ca$ phase also appear. Thus, a few nanoscale intermetallic particles have also nucleated during the SPD process.

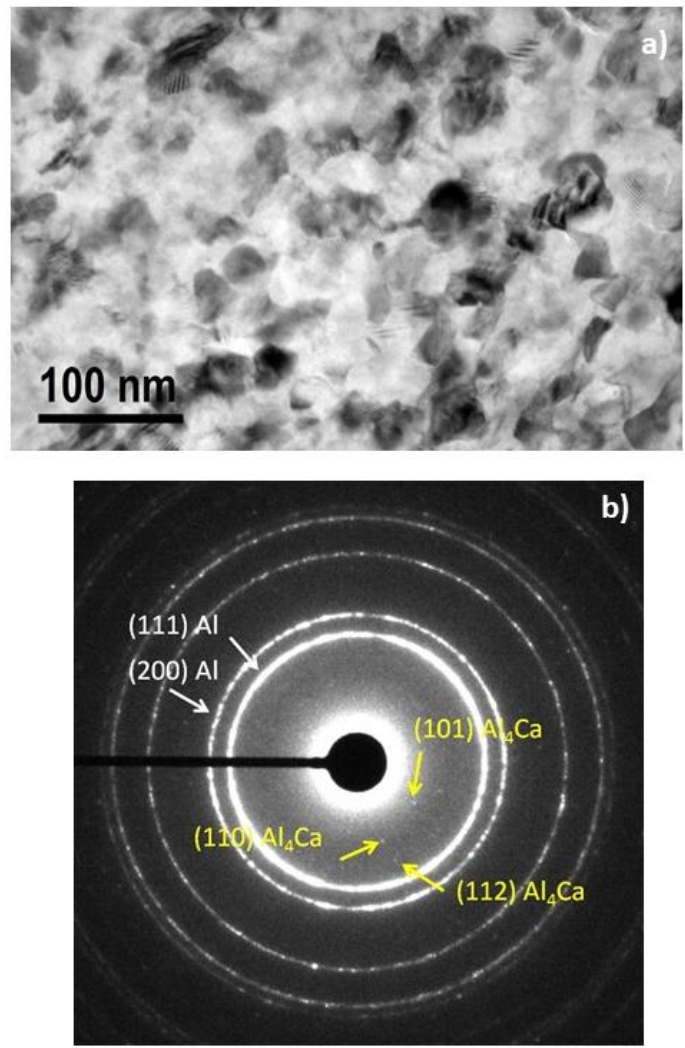

*Figure 4: a) Bright field TEM image of the Al-Ca composite processed by 100 revolutions by HPT showing a nanoscale grain size, b) corresponding SAED pattern exhibiting Debye-Scherrer rings corresponding only to fcc Al. A few additional spots corresponding to the $Al_4Ca$ phase are also exhibited.*

To visualize the distribution of particles within the microstructure, HAADF-STEM imaging was carried out (Fig. 5a). Ca-rich regions are brightly imaged since the atomic number of Ca is much higher than that of Al. The contrast is, however, relatively poor, which is partly due to overlapping nanograins in the TEM sample. The low contrast notwithstanding, it is evident that Ca atoms are not homogeneously distributed. This feature was confirmed using EDS line profiling (Fig. 5b), and it indicates that even if the fcc Ca phase disappeared during SPD, a fully homogeneous super-saturated solid solution was not achieved.

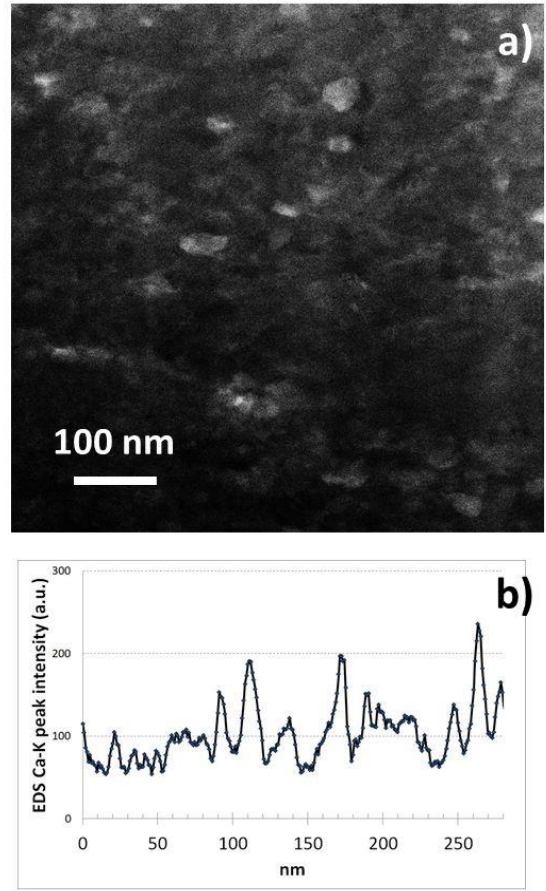

*Figure 5:* STEM dark-field image showing the ultrafine grain structure of the Al-Ca composite processed by 100 revolutions by HPT (a). The slight contrast variations on the HAADF image are attributed to local fluctuations of the Ca content which are confirmed by the EDS line profile analysis (b).

To determine more accurately the distribution of Ca atoms within the nanoscaled structure, the material was analyzed by Atom Probe Tomography (APT). Two representative volumes are displayed in Fig. 6a and 6b. These data confirm that Ca is not homogeneously distributed: there are segregations along linear or planar defects and Ca-rich particles. The composition profile computed across such a nanoparticle (diameter of about 5 nm) shows that it contains about 15 at.% Ca (Fig. 6c). Numerous particles were probed and for all of them, the Ca concentration stands between 10 and 20 at.% Ca.

There are also Ca-depleted regions in reconstructed volumes, which correspond to nanoscale Al grains. The amount of Ca in solid solution was carefully measured, taking care not to include either Ca-rich particles or segregations. The average amount of Ca measured in solid solution was 0.23 at.% (± 0.05), which is much higher than the equilibrium solubility limit at room temperature [38, 39].

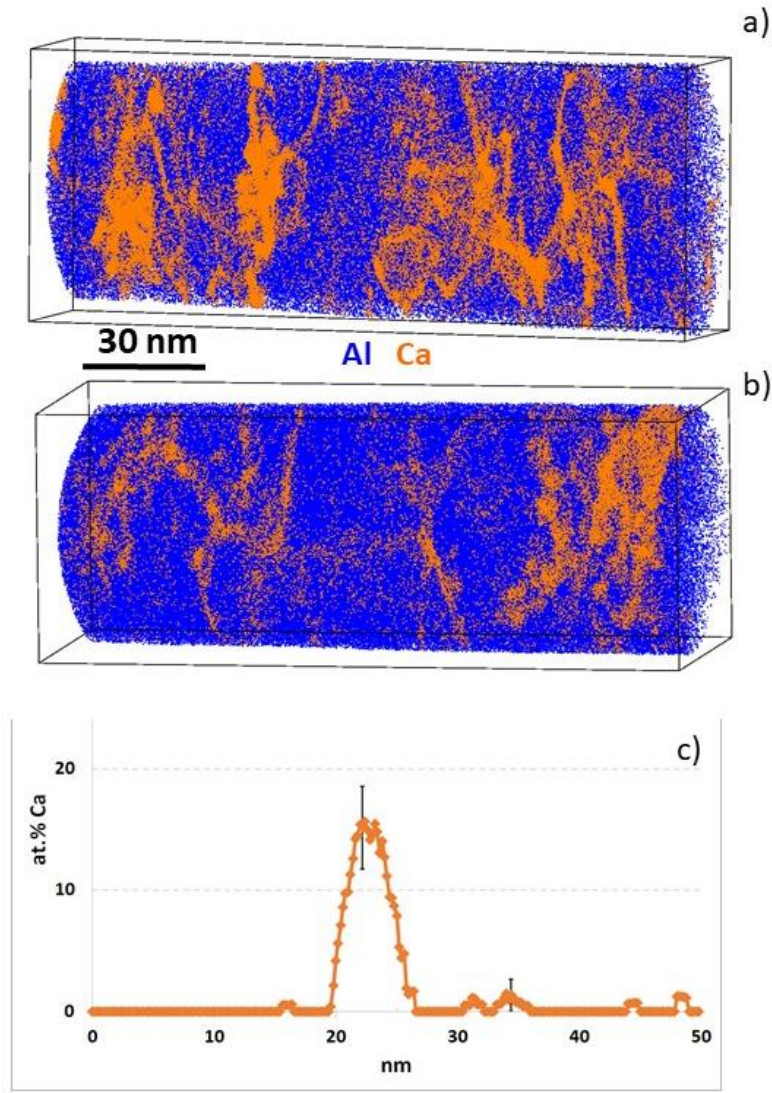

**Figure 6:** (a) and (b) 3D reconstructions of volumes analyzed by APT in the Al-Ca composite processed by 100 revolutions by HPT (Al in blue, Ca in orange). (c) Concentration profile plotted across a Ca-rich nanoscale particle (sampling box thickness 1 nm).

The thermal stability of this nanoscale structure achieved by SPD was investigated by in-situ TEM annealing. On Fig. 7, the same region of the sample that was tracked during annealing at 200 °C is shown in as-SPD condition and after 30 min at 200 °C in STEM-BF images (Fig. 7a and 7c). Comparing these images clearly shows significant grain growth, and STEM-HAADF images indicate nucleation of Ca-rich particles (top left in Fig. 7d) and the coarsening of Ca-rich regions (bottom right in Fig. 7d). Thus, a significant redistribution of Ca atoms occurred during annealing. The evolution of the grain size was estimated from bright-field TEM images (Figs. 4a and 8a) and corresponding distributions are plotted on Fig. 8c. The mean grain size increases from about 25±5 nm to 35±5 nm during 30 min at 200 °C. In addition, the SAED pattern recorded in the annealed state (Fig. 8b) clearly exhibits many spots attributed to the $Al_4Ca$ phase, confirming the nucleation of this Ca-rich phase in agreement with XRD data (Fig. 3).

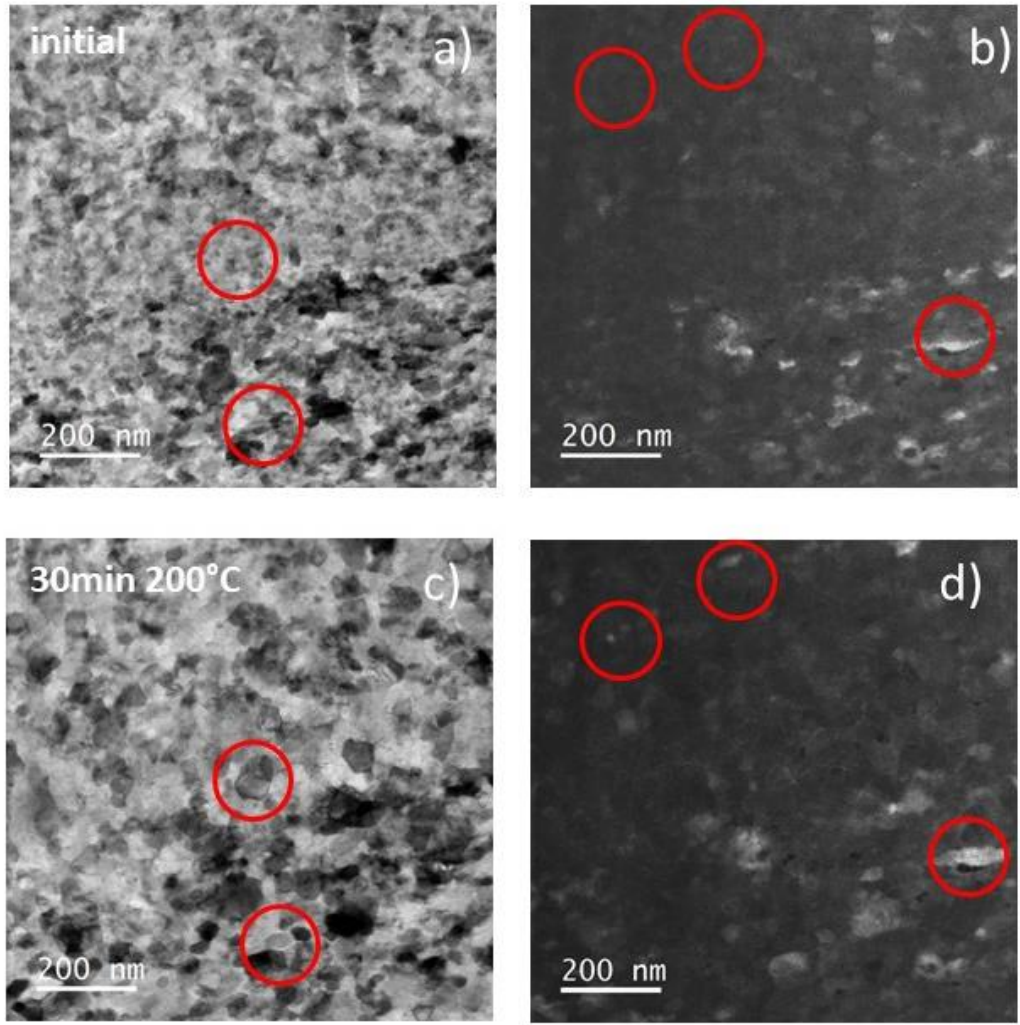

*Figure 7*: *STEM BF (a, c) and HAADF (b, d) images of the nanoscale structure of the Al-Ca composite processed by 100 revolutions by HPT showing the microstructure evolution during in-situ annealing. The same regions are displayed in the initial state (a, b) and after 30 min at 200 °C (c, d). Grain growth and precipitation of Ca-rich clusters are highlighted by the red circles.*

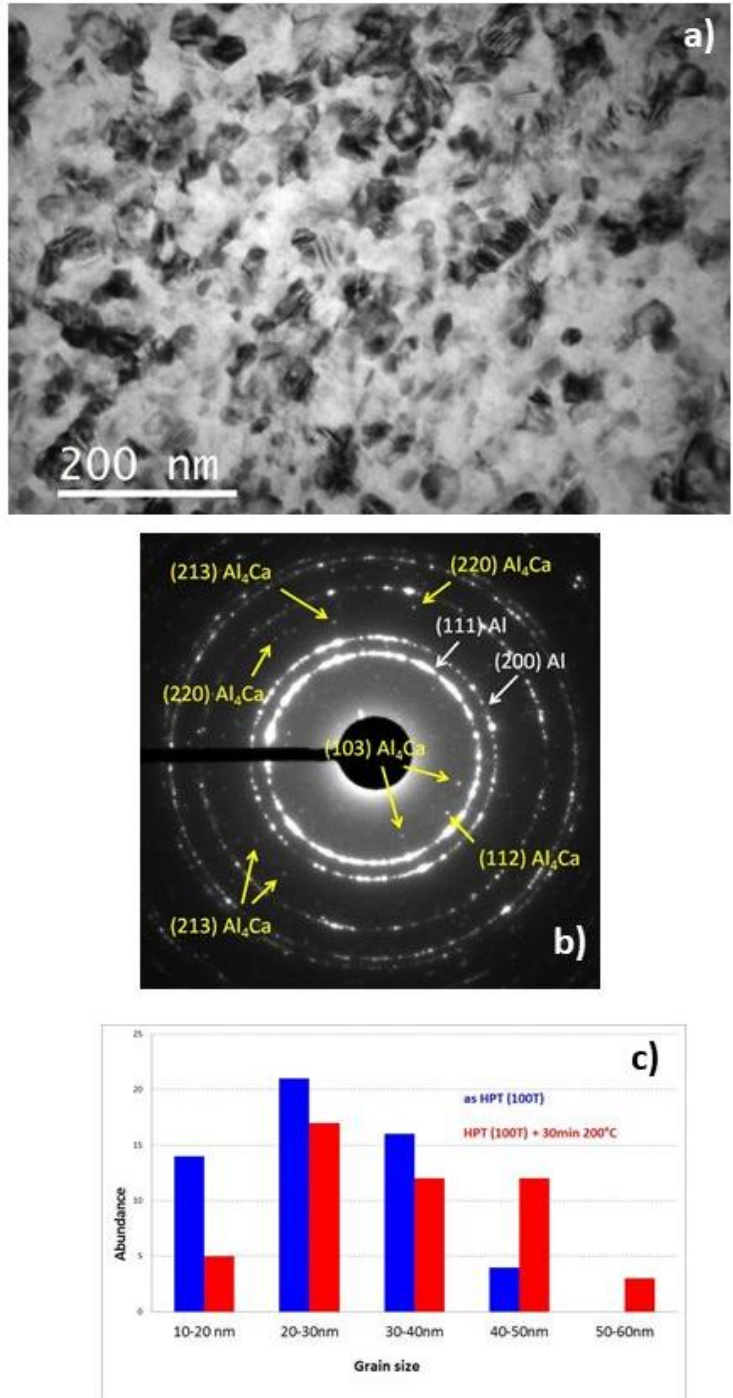

*Figure 8*: *(a) TEM BF image showing the nanoscaled structure after in-situ annealing 30 min at 200 °C and (b) corresponding SAED pattern showing diffraction spots attributed to fcc Al and Al$_4$Ca phase. (c) Grain size distribution measured after 30 min at 200 °C and compared to the initial state.*

The Ca redistribution within the nanoscaled structure was studied more accurately using APT, and a typical volume analyzed in the annealed sample is displayed on Fig. 9a. There are two kinds of Ca-rich particles, some are faceted with a size close to the measured grain size (20-40 nm) and others are much smaller with a diameter in a range of 5 to 10 nm. Composition profiles (Fig. 9b and 9c) show that they all contain about 20 at.% Ca, as expected for the $Al_4Ca$ phase. The average amount of Ca left in solid solution was also measured from these APT data. It is much lower than the as-SPD state, down to only 0.045 at.% (± 0.01), which is closer to the solubility limit [38, 39].

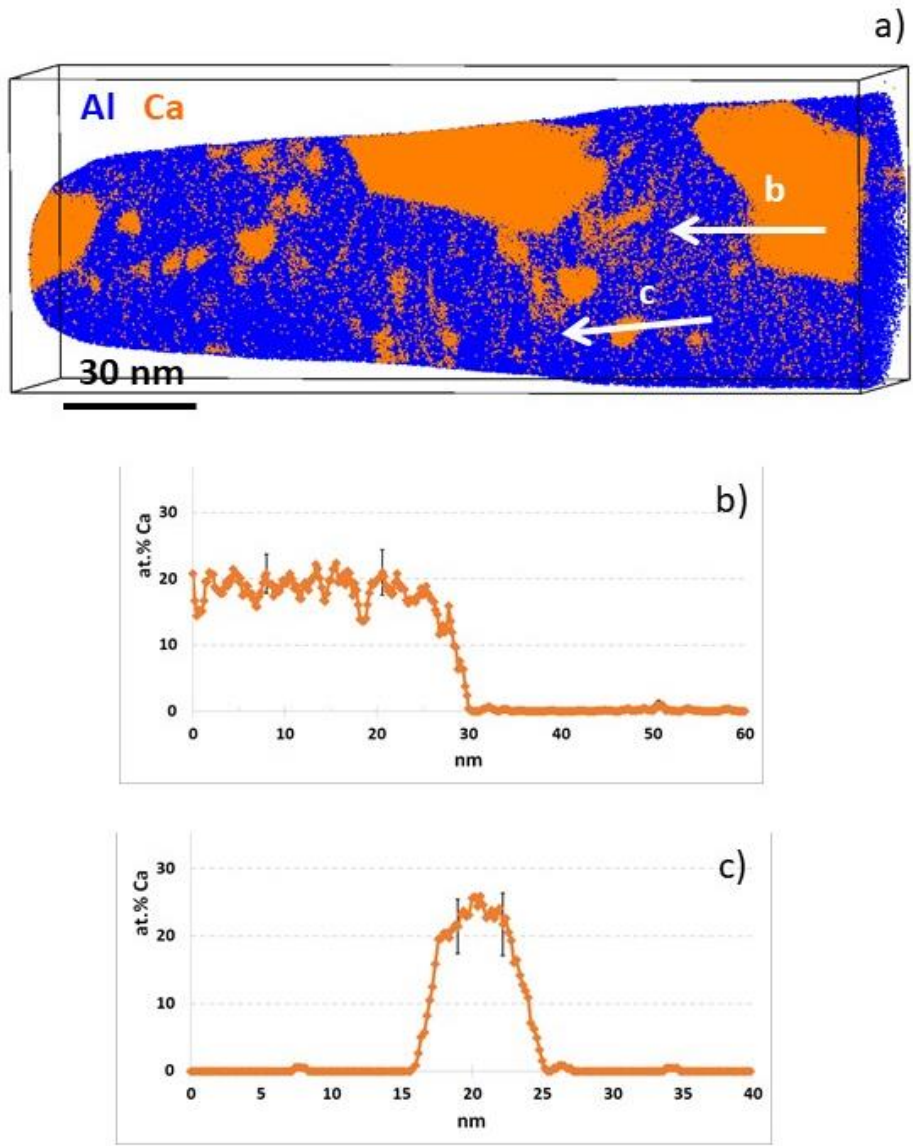

*Figure 9*: *(a) 3D reconstruction of a volume analyzed by APT in the Al-Ca composite processed by 100 revolutions by HPT and annealed during 30 min at 200 °C (Al in blue, Ca in orange). The arrows indicate the location and the direction of the Ca concentration profiles plotted in (b) and (c) (sampling box thickness 1 nm).*

## 4- Discussion

*4.1 Influence of calcium on the deformation induced nanoscaled structure*

In the Al-Ca composite before SPD, there are two phases: the fcc Al matrix and elongated sub-micrometer fcc Ca filaments [37]. These two phases are ductile and easily co-deform during the HPT process, which is the main difference with as-cast structures containing brittle intermetallic particles [19, 28-30]. During the early stage of deformation (one revolution by HPT), only a smooth hardness gradient appears (Fig. 1), although the shear strain gradient is huge across the HPT disc diameter. This is obviously the result of the relatively easy co-deformation between the two phases. At a shear strain of about $\gamma$ = 200, a dramatic change both in electrical resistivity and microhardness was observed (Fig. 2). At this stage, the original sub-micrometer Ca filaments ($d_{Ca}$ < 1µm) are greatly elongated with a thickness $d_{Ca}^* \sim d_{Ca} / \gamma$ of only 5 nm or less. Thus, at larger levels of deformation, Ca filaments are so much elongated that they start to dissolve in the fcc Al matrix as observed in other immiscible systems [31-36]. It starts at the HPT disc periphery where the shear strain is larger, and it progressively reaches the whole sample, leading to a progressive decrease and vanishing of Ca peaks on XRD spectra (Fig. 3). Then, even if the distribution of Ca is not homogeneous within the nanoscaled structure, a kind of saturation regime seems to be reached. As shown by our TEM and APT data (Figs. 4, 5 and 6), a nanoscale structure with an extremely small grain size, much smaller than that achieved by SPD in commercially pure Al, is achieved [23, 14, 15].

Ca solute atoms obviously play a critical role thanks to their strong interaction with crystalline defects, as in the Fe-C system [34]. The distribution of Ca is heterogeneous at different length scales: sub-millimeter, sub-micrometer, and nanoscale. At the larger scales, the non-uniform Ca distribution results mainly from the original distribution of Ca filaments in the composite [37], and it is illustrated by the deviation between the nominal Ca concentration and the measurements done by TEM-EDS and APT (see Table 1). Then at the nanometer scale, due to the extremely low equilibrium solubility in Al [38, 39], only a very small fraction of Ca atoms is driven into solid solution (less than 1 Ca atom out of 10), while others are segregated along crystalline defects such as GBs or located in Ca-rich particles (Fig. 6). The fraction of available sites at GBs $f$ writes as:

$$f \sim \frac{3a}{2d} \qquad (1)$$

Where $a$ is the lattice parameter of fcc Al ($a$ = 0.405 nm) and $d$ the mean grain size (25 nm in the as-deformed material). It yields to $f \sim$ 2% of lattice sites located at GBs, i.e. only twice less than the nominal atomic fraction of Ca and relatively close to the local concentration measured in APT volumes (table 1). Then crystalline defects, and especially GBs become easy nucleation sites for $Al_4Ca$ precipitates and considering the highly negative enthalpy of formation of this intermetallic phase (-17.2 kJ/mol [38]), it is not surprising that few particles directly nucleate along these defects during SPD, as shown on Fig. 6 and confirmed by the SAED pattern (Fig. 4b). It is however interesting to note that their composition estimated from APT data is significantly lower than the expected stoichiometry (in a range of 10 to 20 at.% Ca, Fig. 6c). Nevertheless, they are only a few nanometers in size, and they exhibit an evaporation field slightly lower than the Al matrix (estimated from changes of the local density in reconstructed volumes), thus ion trajectory overlaps may lead to an under-estimation of the Ca content. However, the $Al_4Ca$ stoichiometry was measured in nanoscale particles after annealing (Fig. 9c), thus it could be that in as-SPD samples, particles are highly defected, resulting in a non-stoichiometric composition. Similar features have been reported for $Fe_3C$ carbides in severely deformed steels [40].

When the deformation is continued in the saturation regime ($\gamma \sim 1000$), the $Al_2Ca$ phase nucleates (Fig. 3b). This is not expected from the phase diagram in an alloy containing only 4 at.% Ca [38, 39], but locally the amount of Ca segregated on crystalline defects might be much higher and the nucleation of this phase could be favored due to a significantly lower enthalpy of formation as compared to $Al_4Ca$ (-29.3 kJ/mol for $Al_2Ca$ against -17.2 kJ/mol for $Al_4Ca$ [38]). For such extreme deformations, the formation of metastable phases has been reported in other systems [41]. Both, the high pressure applied during HPT (about 6 GPa) and also possible local temperature increase [42] may also affect the thermodynamic equilibrium. In the Al-Ca composite deformed up to ($\gamma \sim 1000$, it leads to a significant decrease in hardness, probably because this phase being relatively rich in Ca (33 at.%), the amount of Ca available to interact with crystalline defects is then significantly reduced.

Nanoscale structures are intrinsically thermally unstable due to the extremely high fraction of interfaces. In the present case, the low mutual solubility of Ca and Al is another driving force for the transformation of the system as soon as enough atomic mobility is provided to evolve towards a situation with a lower enthalpy. Crystalline defects, such as dislocations and GBs, are fast diffusion paths for the solute, thus in only 30 min at 200 °C, the nanoscale structure has completely changed. Grain growth (Fig. 7 and 8c) and precipitation of the $Al_4Ca$ phase occur concomitantly. If it is assumed that all Ca atoms are located in $Al_4Ca$ particles, then the volume fraction of this phase estimated from the nominal composition is about 24 vol.%. APT data revealed (Fig. 9) that precipitation occurred through two different mechanisms: i) some nanoscale (size below 10 nm) and relatively spherical $Al_4Ca$ particles nucleated homogeneously, inside Al grains, probably directly from the super-saturated solid solution, and eventually fed by pipe diffusion along dislocations; ii) others, larger (20-40 nm) and faceted like typical grains, nucleated at GBs and then grew at a faster rate. This second mechanism is often dominant in the decomposition of UFG super-saturated solid solutions where it is usually impossible to achieve homogeneous precipitation within nanoscale or submicrometer grains [16, 43-45].

## 4.2 Influence of the calcium distribution on the mechanical strength

At low levels of deformation, the easy co-deformation of Al and Ca does not lead to a significant increase in hardness; it remains below 40 HV (Fig. 1). At higher plastic strain, the Ca fcc phase dissolves, and the resulting nanoscale structure (mean grain size of 25±5 nm) gives rise to a huge increase in hardness to about 280-300 HV. The works of H.J. Choi [7] and S.I. Ahmed [46] on nanocrystalline aluminum demonstrate that Tabor's law ($\sigma_y$ = HV (MPa) / 3 ; [47]) provides a good estimate of the yield stress $\sigma_y$ in such materials. Then, following this approach, the corresponding yield stress of the Al-Ca composite processed by SPD was compared to other nanoscale Al alloys on the Hall Petch plot on Fig. 10.

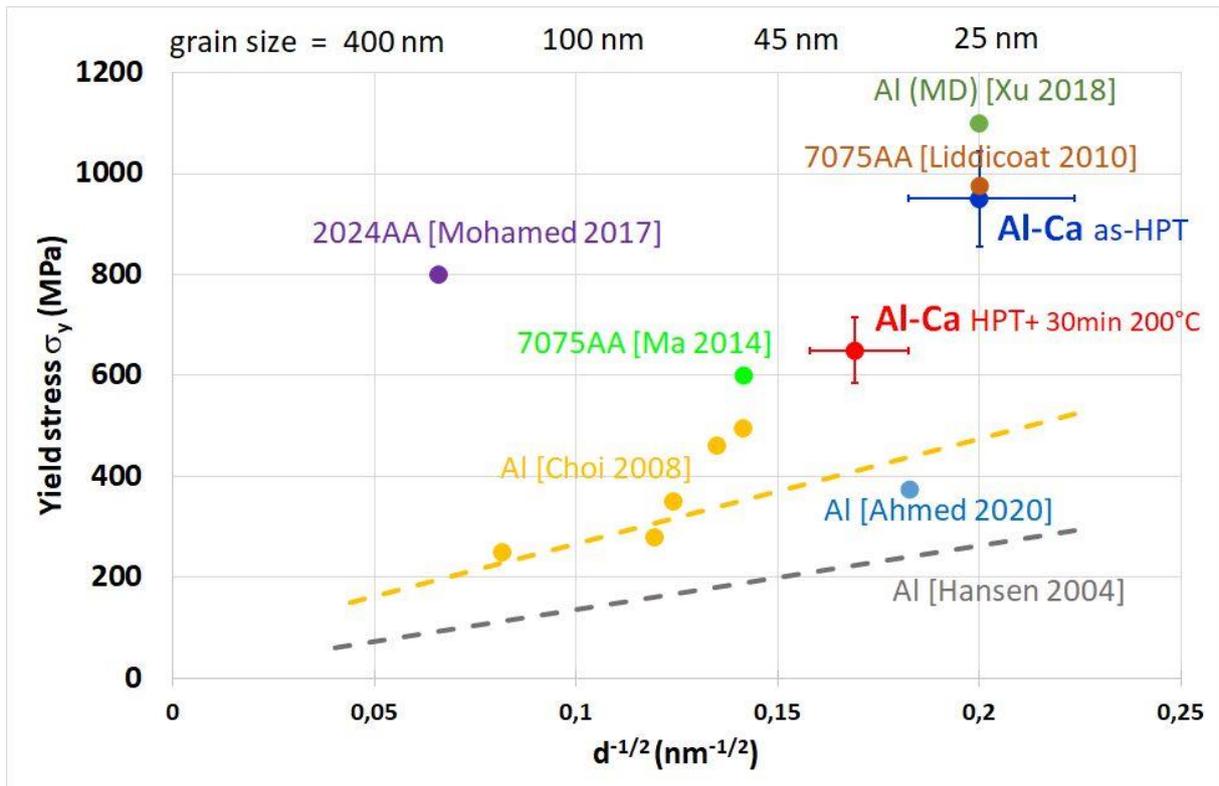

*Figure 10*: Hall-Petch plot (Yield stress as a function of $d^{-1/2}$) comparing the yield stress of the nanostructured Al-Ca composites of the present study (estimated from Tabor's law) with data from the literature. These are all experimental data except for [XU 2018], which refers to molecular dynamic simulations at high strain rate [6]. Data from [Liddicoat 2010] and [Mohamed 2017] were obtained respectively on 7075 and 2024 Al alloys processed by HPT [18, 48]. Data from [Ma 2014] were obtained on a 7075 Al alloy processed by cryo-milling [49]. Data from [Choi 2008] and [Ahmed 2020] were obtained on consolidated ball milled pure Al [7, 46]. The grey dashed line was plotted using the HP coefficient reported by [Hansen 2004] in [8].

For most of the data on this plot, a significant deviation to the classical Hall Petch law ($\sigma_y = \sigma_0 + k\, d^{-1/2}$ where $\sigma_0$ and k are constants) is exhibited. The experimental work of H.J. Choi [7] seems to indicate that such a deviation occurs for a grain size below 80 nm in nanocrystalline Al. At such small grain size, dislocation pile-ups are quite unlikely, and other deformation mechanisms are involved [9-11, 50, 51]. Besides, it is also well accepted that grain boundary mediated plasticity may play a critical role at such length scale [12, 50, 52], especially at relatively low strain rates. Then, it is not surprising that molecular dynamics simulation results [6] with a strain rate much higher than that of experimental data ($10^{10}$ s$^{-1}$ against $10^{-4}$ or $10^{-3}$ s$^{-1}$), exhibits the higher yield stress for a grain size of 25 nm (~ 1.1 GPa [6], against ~ 0.3 GPa predicted by the Hall-Petch law [8]). It is important to note that all experimental data on this plot with a yield stress larger than 600 MPa were obtained on alloys where solute elements, clusters and precipitates are thought to provide additional strengthening contributions. The nanocrystalline Al-Ca alloy (as processed by HPT), with an estimated yield stress of 915-980 MPa is close to the highest reported value of 1 GPa for an ultrafine 7000-series aluminum alloy [18]. There is however only a small amount of Ca in solid solution (about 0.2 at.%, see table1) and a low number density of nanoscaled Ca rich particles mostly located at GBs, thus most of the hardening obviously results from calcium segregations along boundaries. At such small grain sizes, dislocation pile-ups are quite unlikely and grain boundary mediated plasticity probably plays a critical role [12, 50, 52]. Ca atoms located at GBs probably affect the mobility of intergranular dislocations strongly and thus limit grain boundary plasticity which gives rise to the exceptionally high microhardness. Similar effects have been reported in other nanoscaled structures where it was demonstrated that composition modulations or segregations influence the thermal stability but also significantly increase the mechanical strength [52, 53].

After annealing at 200 °C for 30 min, the hardness decreases by about 30%, down to about 200 HV (Fig. 2) as a result of microstructural changes. The mean grain size increased from $d$ = 25±5 nm to 35±5 nm (Fig. 8) and intensive precipitation of the intermetallic Al$_4$Ca phase occurred with a volume fraction of about 24% (Fig. 9). Such a nanoscaled structure is relatively similar to that obtained by Rogachev et al [30] who processed by SPD an as-cast alloy containing intermetallic particles, leading to a similar microhardness. The hardness of Al$_4$Ca being higher than that of Al [54] and the classical Hall-Petch dependence of the yield stress being a function of $d^{-1/2}$, then the expected hardness drop during annealing of the nanostructured Al-Ca material of the present study should not exceed 15%. The significant difference with our experimental observations (30%) clearly indicates that Ca segregations probably inhibit partly grain boundary mediated plasticity leading to the exceptional yield stress close to 1 GPa.

*4.3 Influence of the calcium distribution on electrical properties*

Once the level of deformation becomes moderate (below $\gamma$ = 200), the electrical resistivity remains relatively stable (Fig. 2) with only a minor increase as compared to the initial value of about 300 $10^{-4}\mu\Omega$m [37]. Then, at higher shear strains, when the fcc Ca phase starts dissolving, the electrical resistivity changes dramatically and becomes larger than 2000 $10^{-4}\mu\Omega$m (Fig.2). The electrical resistivity of Al is affected by crystalline defects (vacancies, dislocations and GBs) and Ca in solid solution. For the as-processed nanoscaled Al-Ca material, the influence of nanoscaled Al$_4$Ca particles will be at first neglected since the volume fraction is relatively small, and the contribution of Al/Al$_4$Ca interfaces will be discussed later on. Thus, the electrical resistivity $\rho_{Al}$ can be simply written using the so-called Mattissen-Flemming rule [55]:

$$\rho_{Al} = \rho_{Al}^0 + C_v \Delta\rho_{Al}^v + L_d \Delta\rho_{Al}^d + S_{GB} \Delta\rho_{Al}^{GB} + C_{Ca} \Delta\rho_{Al}^{Ca} \qquad (2)$$

Where $\rho_{Al}^0$ is the electrical resistivity of pure Al (265 $10^{-4}\mu\Omega$m), $C_v$ the vacancy concentration, $L_d$ the dislocation density (m$^{-2}$), $S_{GB}$ the surface of GBs per unit volume (m$^{-1}$), $C_{Ca}$ the fraction of Ca in solid solution, and $\Delta\rho_{Al}^v$, $\Delta\rho_{Al}^d$, $\Delta\rho_{Al}^{GB}$ and $\Delta\rho_{Al}^{Ca}$ the corresponding constants reflecting each contribution (respective values are listed in Table 2).

As shown by Y. Miyajima [23], even if the vacancy concentration could be up to $10^{-3}$% in severely deformed alloys, their contribution to the total electrical resistivity remains several orders of magnitude lower than $\rho_{Al}^0$. The contribution of dislocations could be significantly higher than vacancies and rises up to about 10% of $\rho_{Al}^0$ ($\sim$ 27 $10^{-4}\mu\Omega$m) for a dislocation density of $10^{16}$ m$^{-2}$ (corresponding to a mean distance between dislocations of about 10 nm) [23, 59]. Considering the amount of Ca in solid solution measured by APT (<0.5 at.%, see Table 1), then the corresponding increase in electrical resistivity is less than 15 $10^{-4}\mu\Omega$m. Lastly, grain boundaries, assuming spherical grains with a mean diameter of 25 nm, gives a maximum electrical contribution to the overall resistivity of 300 $10^{-4}\mu\Omega$m. These estimates are listed in Table 2. They lead to an upper theoretical limit of the electrical resistivity of about 610 $10^{-4}\mu\Omega$m, which is significantly lower than the experimental value of 2100±100 $10^{-4}\mu\Omega$m (Fig.2).

It should be noted however that the above approach completely neglects the contribution of Ca atoms segregated along GBs. It has been shown that the grain boundary scattering writes as [60, 61]:

$$\rho^{GB} = \frac{\rho_{Al}^0}{3\left(\frac{1}{3}-\frac{\alpha}{2}+\alpha^2-\alpha^3 ln\left(1+\frac{1}{\alpha}\right)\right)} \qquad (3)$$

With

$$\alpha = \frac{\lambda R}{d (1-R)} \qquad (4)$$

Where $\lambda$ is the mean free path of electrons inside Al grains ($\lambda$ = 18.9 nm [62]), and $R$ the reflection coefficient of grain boundaries ($R$ = 0 means that GBs do not scatter any electron and thus have no influence on the electrical resistivity). H. Schwarz and R. Lück have shown that in metals the value of $R$ is often close to 0.5 [63]. On Fig. 11, the grain boundary contribution to resistivity estimated from eq. (3) was plotted as a function of the grain size for various values of $R$. The comparison with the experimental dependence as reported by Andrews [57] and Kasen [58] indicates that 30% < $R$ < 50% in pure Al. To fit our experimental measurements of electrical resistivity in the nanostructured AlCa, a GB contribution of about 1500 $10^{-4} \mu\Omega$m is required (Table 3). Then, the plot on Fig. 11 also reveals that in this material, the fraction of electrons reflected or trapped by GBs is much higher than that in pure Al, up to about 85%, and this large difference can only be attributed to Ca atoms segregated at GBs.

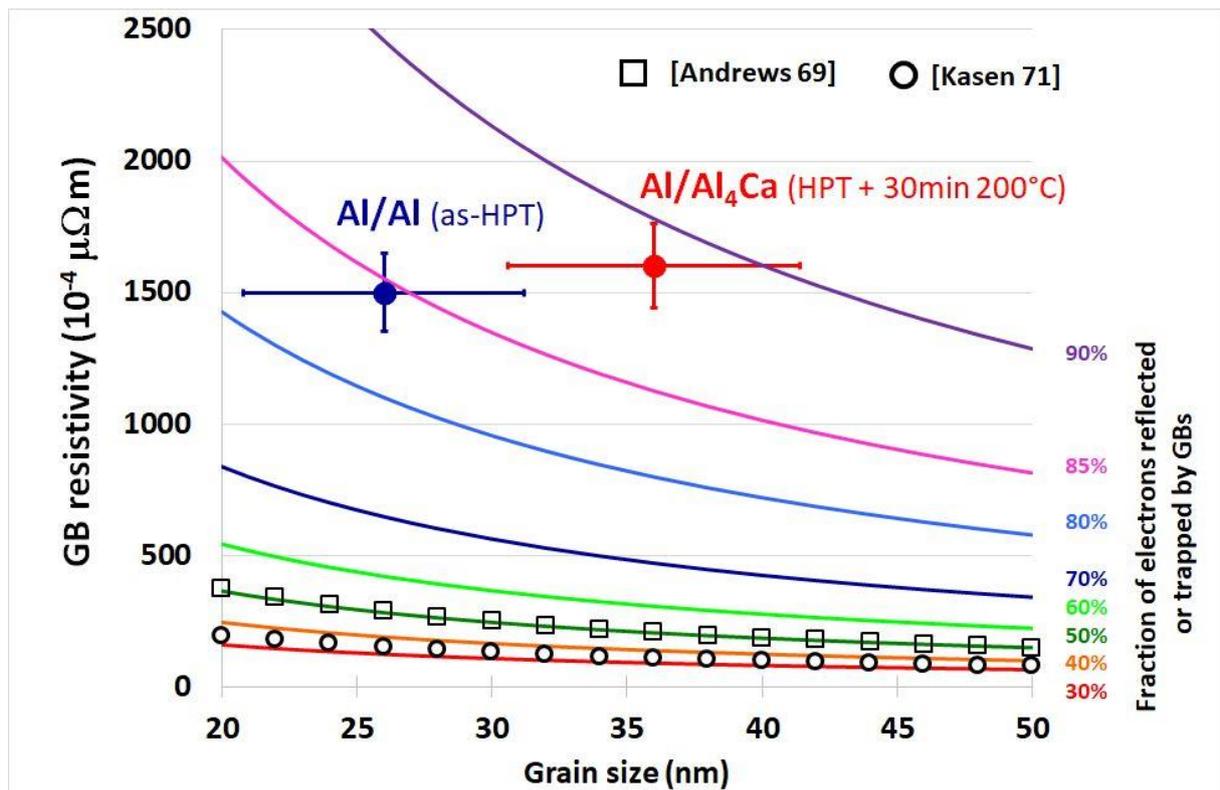

*Figure 11: Contribution of GBs to electrical resistivity of pure Al as a function of the fraction of electrons reflected or trapped by GBs (plotted from eq. (3)). These theoretical data are compared to experimental data of intrinsic GB resistivity of pure Al reported by Andrews [57] and Kasen [58]. The experimentally estimated boundary contributions to electrical resistivity are also indicated for Al GBs in the as-HPT nanostructured alloy and for Al/Al$_4$Ca interfaces in the nanostructured alloy after annealing during 30min at 200°C.*

Annealing at 200 °C during 30 min leads to a strong decrease of the amount of Ca in solid solution in Al (down to 0.045 at.%, see Table 1) and to the precipitation of Al$_4$Ca particles (Fig. 9). Assuming that the nanostructured material is a mixture of pure Al and Al$_4$Ca, then the electrical resistivity can be estimated with a simple rule of mixture [37] :

$$\rho_{Al-Ca} = \rho^0_{Al} f_v^{Al} + \rho^0_{Al4Ca} f_v^{Al4Ca} \qquad (4)$$

Where $\rho_{Al4Ca}{}^0$ is the intrinsic electrical resistivity of the Al$_4$Ca phase (517 10$^{-4}$μΩm [37]), $f_v^{Al}$ and $f_v^{Al4Ca}$ the volume fraction of Al and Al$_4$Ca phases (about 75% and 25% respectively). It leads to an estimated total resistivity after annealing of about 330 10$^{-4}$μΩm. This is much lower than our experimental measurements (about 1200 10$^{-4}$μΩm, Table 3), confirming that at such a small grain size ($d$ = 35±5 nm), electron scattering or trapping by boundaries cannot be neglected. If it is assumed that Al$_4$Ca grains are homogeneously distributed, then only two kinds of interfaces prevail in the system, namely Al/Al and Al/Al$_4$Ca boundaries. Thus, neglecting the contribution of other defects, the electrical resistivity of the annealed nanostructured Al-Ca annealed materials may be written from a combination of eq. (2) and (4):

$$\rho_{Al-Ca}^{annealed} = \rho^0_{Al} f_v^{Al} + \rho^{GB}_{Al} F_{GB}^{Al/Al} + \rho^0_{Al4Ca} f_v^{Al4Ca} + \rho^{GB}_{Al4Ca} F_{GB}^{Al/Al4Ca} \qquad (5)$$

Where $\rho_{Al}{}^{GB}$ is the Al/Al GB contribution to the electrical resistivity (as estimated by eq. (3), with $R$ = 0.5 and $d$ = 35 nm, $\rho_{Al}{}^{GB}$ = 200 10$^{-4}$μΩm). $F_{GB}{}^{Al/Al}$ and $F_{GB}{}^{Al/Al4Ca}$ are the fractions of Al/Al and Al/Al$_4$Ca boundaries respectively. If it is assumed that Al$_4$Ca and Al grains have similar sizes (35 nm), then 25% volume fraction of Al$_4$Ca leads to $F_{GB}{}^{Al/Al}$ = $F_{GB}{}^{Al/Al4Ca}$ = 50%. Then, as summarized in Table 3, eq. (5) gives $\rho_{Al4Ca}{}^{GB}$ ~ 1500 10$^{-4}$μΩm. This contribution is plotted on Fig. 11 and it shows that it corresponds to a fraction of electrons reflected or trapped by Al/Al$_4$Ca boundaries higher than 85%.

5-     **Conclusions**

i)     An Al-Ca composite was severely deformed by HPT at various strain levels. The micro-hardness increased sharply for a shear strain in a range of 100 to 200, and then it saturated at 290±10 HV for larger strain levels. XRD and TEM data showed that the hardness evolution is connected with both the dissolution of the fcc Ca phase and the formation of a nanoscaled structure with a mean grain size of only 25 nm.

ii)     The electrical resistivity followed exactly the same trend, reaching a saturation value of 2100±100 10$^{-4}$μΩm. This is considerably higher than the initial resistivity of about 300 10$^{-4}$μΩm, and it is obviously connected with the mechanical mixing of Ca and Al. APT analyses revealed, however, that Ca is not homogeneously distributed within the Al matrix. Only a small fraction is in solid solution (0.23±0.05 at.%), the rest being segregated along crystalline defects.

iii) The yield stress of this nanoscaled structure, estimated from the Tabor's Law, is in a range of 915 to 980 MPa, much higher than predictions based on the Hall-Petch Law. Due to the relatively low calcium super saturated solution and the low number density of Ca-rich particles, it is mainly attributed to calcium segregations along grain boundaries. The influence of such segregation on plasticity mechanisms will be investigated in the near future.

iv) Annealing for 30 min at 200 °C led to a slight grain growth (mean size 35±5 nm) and intensive nucleation of the $Al_4Ca$ phase. Both homogeneous nucleation of nanoscaled particles (few nanometers in diameter) and heterogenous precipitation (size similar to the Al grains) occurred. APT analyses revealed that the amount of Ca in solid solution decreased to 0.045±0.01 at.%. The annealed nanoscaled structure exhibited a lower microhardness and electrical resistivity than the as-processed material (200±10 HV and 1200±100 $10^{-4}\mu\Omega m$).

v) The analysis of the electrical resistivity data indicated that the proportion of electrons reflected or trapped by GBs is in range of only 30 to 50% in pure Al, but it rose to about 85% in Al/Ca due to Ca segregations. This is the main contribution to the electrical resistivity measured in the as-deformed state. After annealing, there was no more segregation, but $Al/Al_4Ca$ boundaries significantly affect the properties with a ratio slightly higher, in a range of 85 to 90%.

# Tables

|  | Nominal (12 vol.% Ca) | STEM-EDS (as HPT) | APT (as HPT) Full data | APT (HPT+aged) Full data | APT (as HPT) Solid solution in Al | APT (HPT+aged) Solid solution in Al |
|---|---|---|---|---|---|---|
| at.%Ca | 4 | 2.3 ± 0.5 | 1.2 ± 0.4 | 2.27 ± 0.5 | 0.23 ± 0.05 | 0.045 ± 0.01 |

*Table 1: Ca concentration measured experimentally in the HPT-processed Al-Ca composite (compared to the nominal composition). "HPT+aged" refers to a post HPT aging treatment of 30 min at 200 °C. A systematically lower value was found experimentally for all states and techniques, which is attributed to the heterogeneous distribution of Ca filaments in the starting composite material. The error range given for APT data refers to the scatter from one analysis to another. The last two columns give the concentration of Ca measured with sampling boxes set inside Al grains without overlapping with GBs or Ca rich particles to estimate the amount of Ca in super-saturated solid solution.*

|  | Fcc Al | Vacancies [23, 26] | Dislocations [23, 26, 56] | Ca in sol. sol. [55] | GBs [26, 57, 58] | Total estimated electrical resistivity | Measured electrical resistivity |
|---|---|---|---|---|---|---|---|
| Intrinsic elec. res. | $\rho_{Al}^0$ = 265 (*) | $\Delta\rho_{Al}^v$ = 2.6 ($10^{-9} \Omega m\, at.\%^{-1}$) | $\Delta\rho_{Al}^d$ = 2 ($10^{-19}\, \mu\Omega m^3$) | $\Delta\rho_{Al}^{Ca}$ = 3 ($10^{-5} \Omega m\, at.\%^{-1}$) | $\Delta\rho_{Al}^{GB}$ = 1.3-2.5 ($10^{-16}\, \Omega m^2$) |  |  |
| Defect density | --- | $C_v < 10^{-3}$ % | $L_d < 10^{16}\, m^{-2}$ | $C_{Ca} < 0.5$ at.% | $d$ = 25nm |  |  |
| Estimated contrib. to elec. res. | 265 (*) | < 3 $10^{-2}$ (*) | < 27 (*) | < 15 (*) | < 300 (*) | < 610 (*) | 2100±100 (*) |

*Table 2: Summary of the different contributions to the electrical resistivity of the as-processed Al-Ca material (as-HPT) for a shear strain $\gamma$ > 200. The total estimated electrical resistivity was obtained from eq. (2) and data are compared with experimental values (see Fig. 2). The symbol (*) refers to $10^{-4}\, \mu\Omega m$ unit.*

| $10^{-4}\, \mu\Omega m$ | Fcc Al | Al$_4$Ca | Total estimated from rule of mixture (eq. 5) | Measured | contribution of Al/Al GB (eq. 3) ($R$ = 0.5, $d$ = 35nm) | Resulting Al/Al$_4$Ca GB contribution |
|---|---|---|---|---|---|---|
| Nominal elec. Res. | 265 | 517 [37] |  |  | 220 | 1520 |
| Volume fraction | 75% | 25% |  |  | 50% | 50% |
| Elec.res. | 200 | 130 | 330 | 1200±100 | 110 | 760 |

*Table 3: Summary of the different contributions to the electrical resistivity of the nanostructured Al-Ca material after HPT ($\gamma$ > 200) annealed 30min at 200°C. See text for details.*

**Acknowledgements:**

G. Zaher and I. Lomakin are gratefully acknowledged for the electrical resistivity measurements. Z. Grou is acknowledged for the help in 3D reconstructions of APT data. TEM experiments have been carried out on the GENESIS facility which is supported by the Région Normandie, the Métropole Rouen Normandie, the CNRS via LABEX EMC3 and the French National Research Agency as a part of the program "Investissements d'avenir" with the reference ANR-11-EQPX-0020. The author KE thanks the Light Metals Educational Foundation of Japan for a research fund, and the MEXT, Japan for a Grant-in-Aid for Scientific Research (No. 16H04539). Author AR gratefully acknowledges the support of the U.S.D.O.E. Office of Electricity (Ames Lab contract No. DE-AC02-07CH11358), the Electric Power Research Consortium, and the Ames Laboratory Seed Grant Program for enabling production of the original Al-Ca composite.

**Data availability**

The raw and processed data required to reproduce these findings are available on request to the corresponding author.